%% file: Vehicle_Draft_Conference_v5.tex
\documentclass[conference]{IEEEtran}
 \usepackage{amsmath,amssymb}
 \usepackage{subfigure}
 \usepackage{graphicx,graphics,color,psfrag}
 \usepackage{cite,balance}
 \usepackage{caption}
 \allowdisplaybreaks
 \usepackage{algorithm}
 \usepackage{accents}
 \usepackage{amsthm}
 \usepackage{bm}
 \usepackage{algorithmic}
 \usepackage[english]{babel}
 \usepackage{multirow}
 \usepackage{enumerate}
 \usepackage{cases}
 \usepackage{stfloats}
 \usepackage{dsfont}
 \usepackage{color,soul}
 \usepackage{amsfonts}
 \usepackage{cite,graphicx,amsmath,amssymb}
 \usepackage{subfigure}
 \usepackage{fancyhdr}
 \usepackage{hhline}
 \usepackage{graphicx,graphics}
 \usepackage{array,color}

 \usepackage{amsmath}


\newcounter{parentnumber}

\include{header}

\begin{document}
\textheight = 25.3cm

\title{\huge Sensing   Hidden Vehicles by Exploiting Multi-Path V2V Transmission}
\vspace{-40pt}
\author{\vspace{-3pt}Kaifeng Han$^\dag$, Seung-Woo Ko$^\dag$, Hyukjin Chae$^\ddag$, Byoung-Hoon Kim$^\ddag$, and Kaibin Huang$^\dag$\\
\vspace{-3pt} {\small $\dag$ Dept. of EEE, The  University of  Hong Kong, Hong Kong }\\{\small $\ddag$ LG Electronics, S. Korea}\\ \vspace{-10pt} {\small Email: huangkb@eee.hku.hk}}
\maketitle

\vspace{-20pt}
\begin{abstract}
This paper presents a technology of sensing  hidden vehicles by exploiting  multi-path \emph{vehicle-to-vehicle} (V2V)  communication. This overcomes the limitation of existing RADAR technologies that requires \emph{line-of-sight} (LoS), thereby enabling more intelligent manoeuvre in autonomous driving and improving its safety. The proposed technology relies on transmission of orthogonal waveforms over different antennas at the target (hidden)  vehicle. Even without LoS, the resultant received signal enables the sensing vehicle to detect the position, shape, and  driving direction of the hidden vehicle by jointly analyzing the geometry (AoA/AoD/propagation distance)  of individual propagation path.  The accuracy  of the proposed technique is validated by realistic  simulation  including both highway and rural  scenarios.
\end{abstract}

\section{Introduction}
\vspace{-5pt}

\begin{figure}[t]
\vspace{-20pt}
\centering
\includegraphics[width=6.5cm]{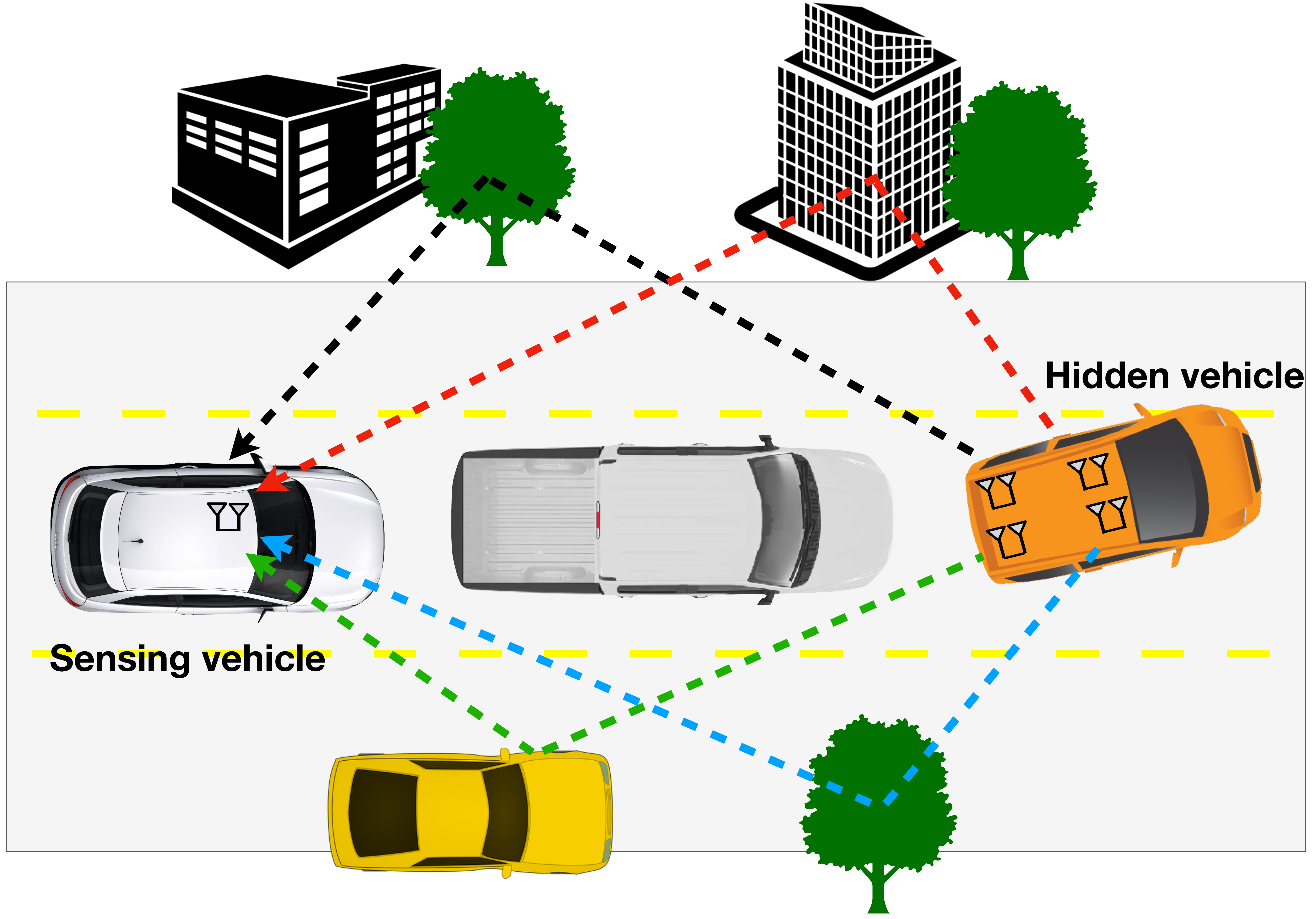}
\caption{Hidden vehicle scenario with multi-path NLoS channels.}\label{v2v_network}
\vspace{-20pt}
\end{figure}

\emph{Autonomous driving} (auto-driving) is a disruptive technology that will reduce car accidents, traffic congestion, and greenhouse gas emissions by automating the  transportation process. One primary operation  of auto-driving is \emph{vehicular positioning}, namely positioning   nearby vehicles and even detecting their shapes \cite{alam2013cooperative}. The positioning includes both absolute and relative positioning and we focus on relative vehicular positioning in this work.
Among others (e.g., cameras and ultrasonic sensing), two  existing technologies, namely RADAR and LiDAR (Light Detection and Ranging),  are capable of accurate  vehicular positioning. RADAR  can localize objects as well as estimate their velocities via sending a designed waveform and analyzing its reflection by the objects. Recent breakthroughs in millimeter-wave radar \cite{choi2016millimeter} or \emph{multiple-input multiple-output} (MIMO) radar \cite{rossi2014spatial} improves the positioning accuracy substaintially. On the other hand,  a LiDAR \cite{schwarz2010lidar} steers  ultra-sharp  laser beams to scan the surrounding environment and generate a high resolution \emph{three-dimensional} (3D) map for  navigation. However, RADAR and LiDAR share the common drawback that positioning requires the target vehicles to be visible with \emph{line-of-sight} (LoS) since neither microwave nor laser beams can penetrate a large solid object such as a truck. Furthermore, hostile weather conditions also affect the effectiveness of LiDAR as fog, snow or rain can severely attenuate a laser beam. On the other hand, detecting hidden vehicles with \emph{non-line-of-sight} (NLoS) is important for intelligent auto-driving (e.g., overtaking) and accidence avoidance  in complex scenarios such as  Fig.~\ref{v2v_network}.

The drawback of existing solutions  motivates the current work on developing a technology for sensing hidden vehicles. It  relies on  V2V transmission to alleviate the severe signal attenuation due to round-trip propagation for RADAR and LiDAR.  By designing hidden vehicle sensing, we aim at tackling two main challenges: 1) the lack of LoS and synchronization between sensing and hidden vehicles and 2) simultaneous detection of position, shape, and orientation of driving direction of hidden  vehicles.

While LiDAR research focuses on mapping, there exist a rich set of signal processing techniques for positioning using  RADAR \cite{gezici2005localization}. They can be largely separated into two themes. The first is  \emph{time-based ranging} by estimating  \emph{time-of-arrival} (ToA) or \emph{time-difference-of-arrival} (TDoA) \cite{shin2002comparisons}. The time-based ranging detects distances but not positions. Most important, the techniques are effective only if there exist LoS paths between sensing and target vehicles.
The second theme  is \emph{positioning using multi-antenna arrays} via detecting  \emph{angle-of-arrival} (AoA) and \emph{angle-of-departure} (AoD) \cite{miao2007positioning}. In addition, there also exist hybrid designs  such as jointly using ToA/AoA/AoD \cite{shahmansoori2018position}.
The techniques make the strong assumption  that perfect synchronization between  transmitters and receivers, which limits their versatility  in auto-driving.
Furthermore, neither time-based ranging nor array-based positioning is capable of additional geometric information on target vehicles such as their shapes and orientation. In summary, prior designs are insufficient for tackling the said challenges which is the objective  of the current work.

The  paper presents a technology for hidden vehicle sensing by exploiting multi-path V2V transmission. The technology requires a hidden vehicle to be provision with an  array with antennas distributed as multiple clusters over the vehicle body. Furthermore, the vehicle transmits a set of orthogonal waveforms over different antennas. Then by analyzing the multi-path signal observed from a receive array, the geometry (AoA/AoD/propagation distance)   of individual path is estimated at the sensing vehicle. Using optimization theory, novel technique is proposed to infer from the multi-path geometric information the position, shape,  and orientation of the hidden vehicles.  Comprehensive simulation is performed  based on  practical vehicular channel model including both highway and rural scenarios. Simulation results show the effectiveness of the proposed technology in sensing hidden vehicles.

\section{System Model}\label{sec:SigModel}
We consider a two-vehicle system where a \emph{sensing vehicle} (SV) attempts to detect the position, shape, and orientation of a \emph{hidden vehicle} (HV) blocked by obstacles such as trucks or buildings (see Fig.~\ref{v2v_network}). For the task of only detecting the position and orientation  (see Section~\ref{sec:singlePositioning}), it is sufficient for HV to have an array of collocated antennas (with negligible half-wavelength spacing). On the other hand, for the task of simultaneous detection of position, shape, and orientation  (see Section~\ref{sec:fourPositioning}), the antennas  at the HV are assumed to be distributed as multiple clusters of collocated antennas over HV body. For simplicity, we consider $4$-cluster arrays with clusters at the vertices of a rectangle. Then sensing reduces to detect the positions and shape of the  rectangle, thereby also yields the orientation of HV. The relevant technique can be easily  extended to  a general arrays topology. Last, the SV is provisioned with a $1$-cluster array.

\subsection{Multi-Path NLoS Channel}\label{subsec:ScenarioDescription}
\begin{figure}[t]
\vspace{-10pt}
\centering
\includegraphics[width=7cm]{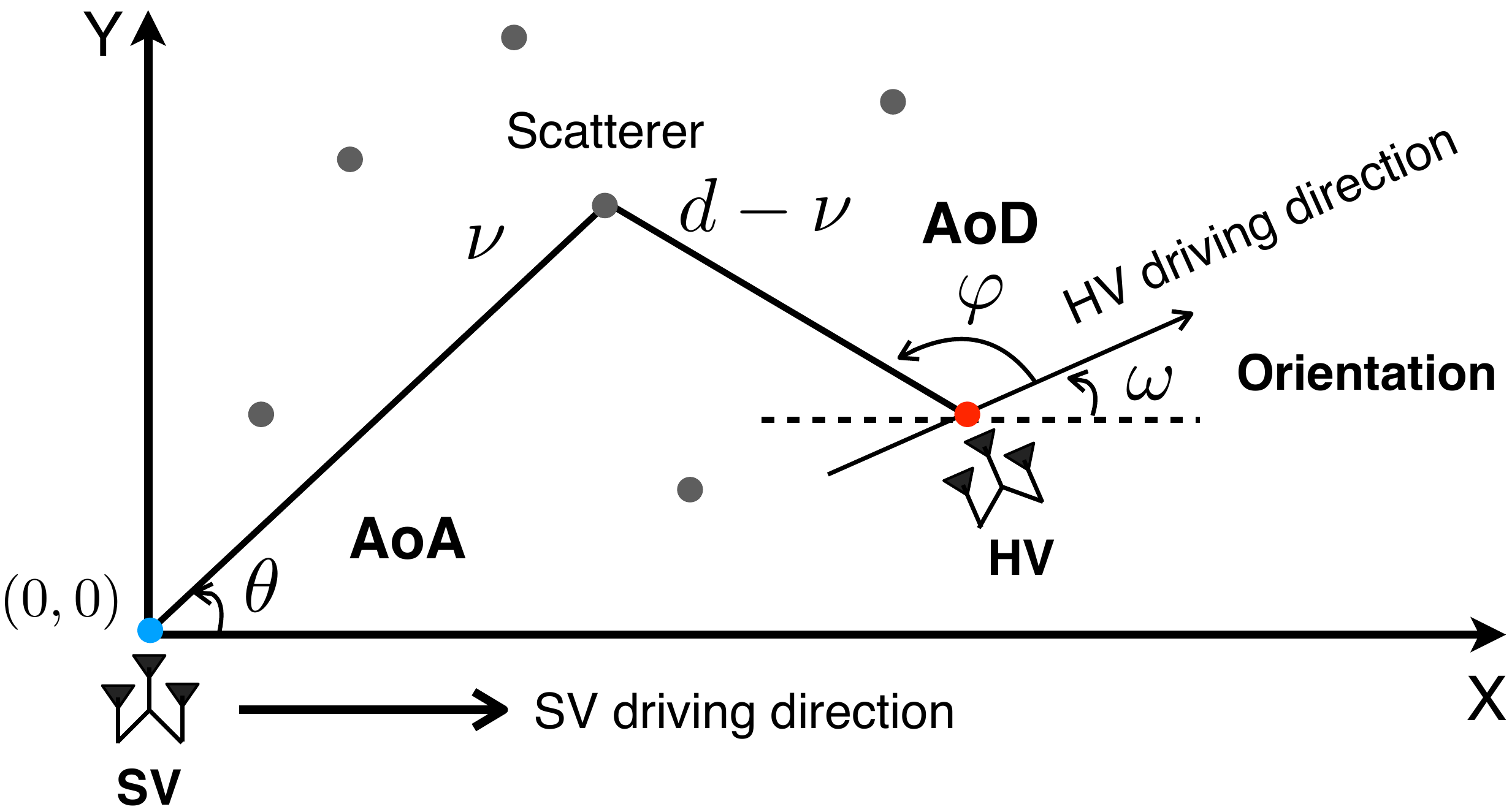}
\caption{NLoS signal model.}\label{2DSignalModel}
\vspace{-20pt}
\end{figure}

The channel between the SV and HV contains NLoS and multi-paths reflected by a set of scatterers. Following the typical assumption for V2V channels, only the received signal from paths with single-reflections is considered at the SV while higher order reflections are neglected due to severe attenuation \cite{Cheng2013}. Propagation is assumed to be constrained within the horizontal plane to simplify exposition. Consider a 2D Cartesian coordinate system where the  SV array  is located at the origin and  the $X$-axis is aligned with the orientation of SV. Consider a typical $1$-cluster  array at the HV.  Each NLoS signal path from the HV antenna cluster   to the SV array
 is characterized by the following five parameters (see Fig.~\ref{2DSignalModel}): the AoA at SV denoted by $\theta$; the AoD at HV denoted by $\varphi$; the orientation of the HV denoted by $\omega$; and the propagation distance denoted by $d$ which includes  the propagation distance before refection, denoted by $\nu$, and the remaining distance $d - \nu$. The AoD and AoA are defined as azimuth angles relative to driving directions of HV and SV, respectively.

\subsection{Hidden Vehicle Transmission}\label{subsec:vehicleTrans}
Each of  $4$-cluster arrays of HV has $M_t$ antennas. The HV is assigned four sets of $M_t$ orthogonal waveforms for transmission. Each set is transmitted using a corresponding  antennas cluster where each antenna transmits an orthogonal waveform. It is assumed that by network coordinated waveform assignment, HV waveform sets are known at the SV that can hence group the signal paths according their originating antennas clusters arrays. Let $s_m(t)$ be the continuous-time baseband waveform assigned to the $m$-th HV antenna with the bandwidth $B_s$. Then the waveform orthogonality is specified by  $\int s_{m_1}(t) s^*_{m_2}(t)dt=\delta(m_1-m_2)$ with the delta  function  $\delta(x)=1$ if $x=0$ and $0$ otherwise. The transmitted waveform vector  for the $k$-th array of HV antennas cluster is  $\mathbf{s}^{(k)}(t) = [s^{(k)}_1(t), \cdots, s^{(k)}_{M_t}(t)]^{\mathrm{T}}$. With the knowledge of $\{\mathbf{s}^{(k)}(t)\}$, the SV with $M_r$ antennas scans  and retrieves the receive signal due to the HV transmission.

Consider a typical HV antennas cluster array.  Based on the far-field propagation model \cite{cui2016vehicle} ,  the cluster response vector     is represented as a function of AoD $\varphi$ as
\begin{align}\label{HVsteeringVector}
\mathbf{a}(\varphi) = [ \exp(j 2\pi f_c \alpha_{1}(\varphi)), \cdots,\exp(j 2\pi f_c \alpha_{M_t}(\varphi))   ]^{\mathrm{T}},
\end{align}
where $f_c$ denotes the carrier frequency and $\alpha_{m}(\varphi)$
refers to the difference in   propagation time to the corresponding scatterer between  the $m$-th HV antenna  and  the $1$-st HV antenna in the same cluster, i.e., $\alpha_{1}(\varphi)=0$.
Similarly, the response  vector of SV array is expressed in terms of AoA  $\theta$ as
\begin{align}\label{SVsteeringVector}
\mathbf{b}(\theta) = [ \exp(j 2\pi f_c \beta_{1}(\theta)), \cdots,\exp(j 2\pi f_c \beta_{M_r}(\theta))   ]^{\mathrm{T}},
\end{align}
where $\beta_{m}(\theta)$ refers to the difference of propagation time from the scatterer to the $m$-th SV antenna than the $1$-st SV antenna. We assume that SV has prior knowledge of the response functions $\mathbf{a}(\varphi)$ and $\mathbf{b}(\theta)$. This is feasible by standardizing  the vehicular arrays' topology.   In addition, the Doppler effect is ignored based on the assumption that the Doppler frequency shift is much smaller than the waveform bandwidth and thus does not affect waveform orthogonality.

Let $k$ with $ 1\leq k\leq 4$ denote the index of HV arrays  and $P^{(k)}$ denote the number of received paths originating  from the $k$-th antennas cluster array. The total number of  paths arriving  at SV is  $P = \sum_{k=1}^4 P^{(k)}$.
Represent  the received signal vector at SV as $\mathbf{r}(t) = [r_1(t), \cdots, r_{M_r}(t)]^{\mathrm{T}}$. It can be  expressed in terms of $\mathbf{s}(t)$, $\mathbf{a}(\varphi)$ and $\mathbf{b}(\theta)$  as
\begin{align}
\mathbf{r}(t) = \sum_{k=1}^{4} \sum_{p=1}^{P^{(k)}} \gamma_{p}^{(k)} \mathbf{b}\l(\theta_{p}^{(k)}\r) \mathbf{a}^{\mathrm{T}}\l(\varphi_{p}^{(k)}\r)\mathbf{s}\l(t - \lambda_p^{(k)} \r) + \mathbf{n}(t), \nn
\end{align}
where $\gamma_{p}^{(k)}$ and  ${\lambda_p^{(k)}}$ respectively denote the complex channel coefficient and
 ToA of path $p$ originating  from the  $k$-th HV array,
and  $\mathbf{n}(t)$ represents channel noise. Without synchronization between  HV and SV,  SV has no information of HV's transmission timing. Therefore, $\lambda_{p}^{(k)}$ differs from the corresponding propagation delay, denoted by
${\tau_p^{(k)}}$, due   unknown clock synchronization gap between HV and SV denoted by $\Gamma$. Consequently, $\tau_{p}^{(k)}=\lambda_{p}^{(k)}-\Gamma$.

\subsection{Estimations of AoA, AoD, and ToA}\label{sec:music}
The sensing techniques in the sequel assume that the SV has the  knowledge of AoA, AoD, and ToA of each receive NLoS signal path, say path $p$, denoted by $\l\{\theta_{p}, \varphi_{p}, \lambda_{p}\r\}$ where $p \in \mathcal{P} = \{1, 2, \cdots, P\}$. The knowledge can be acquired by applying  classical parametric estimation techniques briefly sketched as follows.  The estimation procedure comprises the following three steps. 1) \textbf{Sampling}: The received analog signal $\mathbf{r}(t)$ and the waveform vector $\mathbf{s}(t)$ are sampled at the Nyquist rate $2B_s$ to give discrete-time  signal vectors $\mathbf{r}[n]$ and $\mathbf{s}\l[n\r]$, respectively. 2) \textbf{Matched filtering}: The sequence of $\mathbf{r}[n]$ is matched-filtered using  $\mathbf{s}\l[n\r]$. The resultant $M_r\times M_t$ coefficient matrix $\mathbf{y}[z]$ is given by $\mathbf{y}[z]=\sum_n \mathbf{r}[n]\mathbf{s}^*\l[n - z\r]$. The sequence of ToAs $\{\lambda_{p}\}$ can be estimated by detecting peaks of the norm of $\mathbf{y}[z]$, denoted by $\{z_p\}$, which can be converted into time by multiplying the time resolution $\frac{1}{2B_s}$. 3) \textbf{Estimations of AoA/AoD}: Given $\{\mathbf{y}[z_p]\}$, AoAs and AoDs are jointly estimated using
a 2D-\emph{multiple signal classification} (MUSIC) algorithm \cite{mathews1994eigenstructure}. The estimated AoA $\theta_{p}$, AoD $\varphi_{p}$, ToA $\lambda_{p}$ jointly characterize the $p$-th NLoS path.

\subsection{Hidden Vehicle Sensing Problem}\label{Subsec:TechnicalChallenging}

The SV attempts to sense the HV's position, shape, and orientation.
The position and shape of HV can be obtained by using parameters of AoA $\theta$, AoD $\varphi$, orientation $\omega$, distances $d$ and $\nu$, length and width of configuration of $4$-cluster arrays denoted by $L$ and $W$, respectively. Noting the first two parameters are obtained based on the estimations in Section~\ref{sec:music} and the goal is to estimate the remaining five parameters.

\section{Sensing Hidden Vehicles  with Colocated Antennas}\label{sec:singlePositioning}
Consider the case that the HV has an array with colocated antennas ($1$-cluster array). SV is capable of detecting the HV position, specified by the coordinate $\bp = (x, y)$,  and orientation, specified $\omega$ in Fig.~\ref{2DSignalModel}. The prior knowledge that the SV has for sensing is the parameters of $P$ NLoS paths estimated as described in Section \ref{sec:music}. Each path, say path $p$,  is characterized by the parametric set $\{\theta_p, \phi_p, \lambda_p\}$. Then the \textbf{sensing problem} in the current case can be represented as
\begin{equation}\label{Eq:ProbForm1}
\quad\bigcup_{p \in \mathcal{P}} \{\theta_p, \phi_p, \lambda_p\} \Rightarrow \{\bp, \omega\}.
\end{equation}
The problem is solved in the following subsections.
\subsection{Sensing Feasibility Condition}
In this subsection, it is shown that for the sensing to be feasible, there should exist at least \emph{four} NLoS paths. To this end, based on the path geometry (see Fig. \ref{2DSignalModel}), we can obtain the following system of equations:
\begin{align}\label{FGrelation}\tag{P1}
\begin{cases}
x_{p} = \nu_{p} \cos(\theta_{p}) - (d_{p} - \nu_{p}) \cos(\varphi_{p} + \omega) \\= \nu_{1} \cos(\theta_{1}) - (d_{1} - \nu_{1}) \cos(\varphi_{1} + \omega),  \\
y_{p} = \nu_{p} \sin(\theta_{p}) - (d_{p} - \nu_{p}) \sin(\varphi_{p} + \omega) \\ =\nu_{1} \sin(\theta_{1}) - (d_{1} - \nu_{1}) \sin(\varphi_{1} + \omega),
\end{cases}
 p  \in \mathcal{P}.
\end{align}
The number of  equations in \ref{FGrelation} is $2(P-1)$, and the above system of equations has a unique solution when the dimensions of unknown variables are less than $2(P-1)$.
Since the   AoAs $\{\theta_{p}\}$ and AoDs $\{\varphi_{p}\}$ are known,  the number of unknowns is  $(2P+1)$ including the propagation distances $\{d_{p}\}$, $\{\nu_{p}\}$, and orientation $\omega$. To further reduce the number  of unknowns, we use the propagation time difference between signal paths also known as TDoAs, denoted by $\{\rho_{p}\}$, which can be obtained from the difference of ToAs as
$\rho_{p}=\lambda_{p}-\lambda_{1}$
where   $\rho_{1}=0$.
The propagation distance of signal path $p$, say $d_{p}$, is then expressed in terms of $d_{1}$ and $\rho_{p}$ as
\begin{align}\label{Eq:TDoAandPropagationDist}
d_{p}=c(\lambda_{p}-\Gamma)
=c(\lambda_1-\Gamma)+c(\lambda_{p}-\lambda_{1})=d_{1}+c\rho_{p},
\end{align}
for  $p = \{2, \cdots, P\}$. Substituting the above $(P-1)$ equations into   \ref{FGrelation} eliminates the unknowns $\{d_2, \cdots, d_P\}$ and hence reduces  the number  of unknowns from $(2P+1)$ to $(P+2)$. As a result, \ref{FGrelation} has a unique solution when $2(P-1)\geq P+2$.

\begin{proposition}[Sensing feasibility condition]\label{pro:minNumPath}
\emph{To sense the position and orientation of a HV with $1$-cluster array, at least four NLoS signal paths are required: $P\geq 4$.}
\end{proposition}

\begin{remark}[Asynchronization and TDoA]\emph{Recall that one sensing challenge is asynchronization between HV and SV represented by  $\Gamma$, which is a latent variable we cannot observe explicitly. Considering TDoA helps solve  the  problem by avoiding the need of considering $\Gamma$ by exploiting the fact that all NLoS paths experience the same synchronization gap. }
\end{remark}

\subsection{Hidden Vehicle Sensing  without Noise}\label{Sec:PositionSingleMIMO}
Consider the case of a high receive \emph{signal-to-noise ratio} (SNR) where noise can be neglected. Then the sensing problem in \eqref{Eq:ProbForm1} is translated to solve the system of equations in  \ref{FGrelation}. One challenge is that the unknown orientation $\omega$ introduces  nonlinear relations, namely  $\cos(\varphi_p + \omega)$ and $\sin(\varphi_p + \omega)$, in the equations. To overcome the difficulty, we adopt the following two-step approach: 1) Estimate the correct orientation $\omega^*$ via its discriminant introduced in the sequel;  2) Given $\omega^*$, the equations becomes linear and thus can be solved via  \emph{least-square} (LS) estimator, giving the position $\bp^*$. To this end, the equations in  \ref{FGrelation} can be  arranged in a matrix form as
\begin{align}\label{ax=b}\tag{P2}
\mathbf{A}(\omega) \mathbf{z}= \mathbf{B}(\omega),
\end{align}
where $\mathbf{z}=(\mathbf{v}, d_1)^{\mathrm{T}} \in \mathds{R}^{(P+1)\times 1}$ and  $\mathbf{v} = \{\nu_1, \cdots,   \nu_P\}$. For matrix $\mathbf{A}(\omega)$, we have
\begin{align}\label{matrixA}
\mathbf{A}(\omega) &=
\begin{bmatrix}
  \mathbf{A}^{(\cos)}(\omega)\\
  \mathbf{A}^{(\sin)}(\omega)
\end{bmatrix}
\in \mathds{R}^{2(P-1) \times (P+1)},
\end{align}
where $\mathbf{A}^{(\cos)}(\omega)$ is
\begin{align}\label{A}
\begin{bmatrix}
    a_{1}^{(\cos)} & -a_{2}^{(\cos)}  & 0 & \cdots &0 & a_{1,2}^{(\cos)}  \\
    a_{1}^{(\cos)}   & 0 & -a_{3}^{(\cos)}  & \cdots& 0 & a_{1,3}^{(\cos)} \\
    \vdots  & \vdots & \vdots & \ddots & \vdots & \vdots \\
    a_{1}^{(\cos)}   & 0 & 0 &\cdots &-a_{P}^{(\cos)}  & a_{1,P}^{(\cos)} \\
\end{bmatrix}
\end{align}
with  $a_{p}^{(\cos)}  = \cos(\theta_p) + \cos(\varphi_p + \omega)$ and $a_{1,p}^{(\cos)} =  \cos(\varphi_p +\omega) - \cos(\varphi_1 + \omega)$, and $\mathbf{A}^{(\sin)}(\omega)$  is obtained by replacing all $\cos$ operations in \eqref{A} with $\sin$ operations. Next,
\begin{align}\label{matrixB}
\mathbf{B}(\omega) =
\begin{bmatrix}
  \mathbf{B}^{(\cos)}(\omega)\\
  \mathbf{B}^{(\sin)}(\omega)
\end{bmatrix}
\in \mathds{R}^{2(P-1)\times 1},
\end{align}
where
\begin{align}\label{B}
\mathbf{B}^{(\cos)}(\omega) =
\begin{bmatrix}
    c \rho_2 \cos(\varphi_2 + \omega) \\
    c \rho_3 \cos(\varphi_3 + \omega) \\
    \vdots \\
    c \rho_P  \cos(\varphi_P + \omega) \\
\end{bmatrix},
\end{align}
and $\mathbf{B}^{(\sin)}(\omega)$ is obtained by replacing all $\cos$ in \eqref{B} with $\sin$.

\noindent \underline{1) Computing $\omega^*$:} Note that \ref{ax=b} becomes an \emph{over-determined} linear system of equations if $P \geq 4$ (see Proposition \ref{pro:minNumPath}), providing the following discriminant of orientation $\omega$. Since the equations in \eqref{matrixA} are based on the geometry of multi-path propagation and HV orientation as illustrated in Fig.~\ref{2DSignalModel}, there exists a unique solution for  the equations. Then we can obtain from  \eqref{matrixA} the following  result useful for computing $\omega^*$.

\begin{proposition}[Discriminant of orientation]\label{pro:discriminant}
\emph{With $P \geq 4$, a unique $\omega^*$ exists when $\mathbf{B}(\omega^*)$ is orthogonal to the null column space of $\mathbf{A}(\omega^*)$ denoted by $\mathsf{null}(\mathbf{A}(\omega^*)^{\mathrm{T}})\in \mathds{R}^{2(P-1)\times (P-3)}$:
\begin{align}\label{Eq:OrthogonalNullSpace}
\mathsf{null}(\mathbf{A}(\omega^*)^{\mathrm{T}})^{\mathrm{T}}\mathbf{B}(\omega^*)=\mathbf{0}.
\end{align}}
\end{proposition}
Given  this discriminant, a simple 1D search can be performed over the range $[0, 2\pi]$ to find  $\omega^*$.

\noindent \underline{2) Computing $\bp^*$:}
Given the $\omega^*$, \ref{ax=b} can be  solved by
\begin{align}\label{LSresult}
\mathbf{z}^* = \l[ \mathbf{A}(\omega^*)^{\mathrm{T}} \mathbf{A}(\omega^*)  \r]^{-1} \mathbf{A}(\omega^*)^{\mathrm{T}} \mathbf{B}(\omega^*).
\end{align}
Then  the estimated HV position   $\bp^{*}$ can be  computed  by substituting \eqref{Eq:OrthogonalNullSpace} and \eqref{LSresult} into \eqref{Eq:TDoAandPropagationDist} and \ref{FGrelation}.


\subsection{Hidden Vehicle Sensing  with Noise}\label{subsection:NoiseHVSensing}
In the presence of significant channel noise,  the estimated  AoAs/AoDs/ToAs contain errors. Consequently, HV sensing is based on  the noisy versions of matrix $\mathbf{A}(\omega)$ and $\mathbf{B}(\omega)$, denoted by $\tilde{\mathbf{A}}(\omega)$ and $\tilde{\mathbf{B}}(\omega)$, which do not satisfy the  equations  in \ref{ax=b} and \eqref{Eq:OrthogonalNullSpace}. To overcome the difficulty, we develop a sensing technique by converting the equations into   minimization problems whose solutions are robust against noise.

\noindent \underline{1) Computing $\omega^*$:} Based on \eqref{Eq:OrthogonalNullSpace}, we formulate  the following problem for finding  the orientation $\omega$:
\begin{align}\label{NullSearching_Noisy}
\omega^* = \arg\min_{\omega} \l[\mathsf{null}(\tilde{\mathbf{A}}(\omega)^{\mathrm{T}})^{\mathrm{T}}\tilde{\mathbf{B}}(\omega)\r].
\end{align}
Solving the problem relies on a 1D search over $[0, 2\pi]$.

\noindent \underline{2) Computing $\bp^*$:} Next, given $\omega^*$, the optimal $\mathbf{z}^*$ can be derived by
using the  LS estimator that minimizes the squared Euclidean distance as
\begin{align}\label{LSEstimator_Noisy}
\mathbf{z}^*  &= \arg \min_{\mathbf{z}} \| \tilde{\mathbf{A}}(\omega^*) \mathbf{z} - \tilde{\mathbf{B}}(\omega^*) \|^2\nonumber\\
&=\l[ \tilde{\mathbf{A}}(\omega^*)^{\mathrm{T}} \tilde{\mathbf{A}}(\omega^*)  \r]^{-1} \tilde{\mathbf{A}}(\omega^*)^{\mathrm{T}} \tilde{\mathbf{B}}(\omega^*),
\end{align}
which has the same structure as \eqref{LSresult}.  Last, the origins of all paths $\{(x_p, y_p)\}_{p \in \mathcal{P}}$ can be  computed  using the parameters $\{\mathbf{z}^* , \omega^*\}$ as illustrated in \ref{FGrelation}. Averaging these origins gives the estimate of the HV position $\bp^*=(x^*, y^*)$ with $x^* = \frac{1}{P}\sum_{p=1}^{P} x_p$ and $y^* = \frac{1}{P} \sum_{p=1}^{P} y_p$.


\begin{figure}[t]
\vspace{-20pt}
\centering
\includegraphics[width=6.5cm]{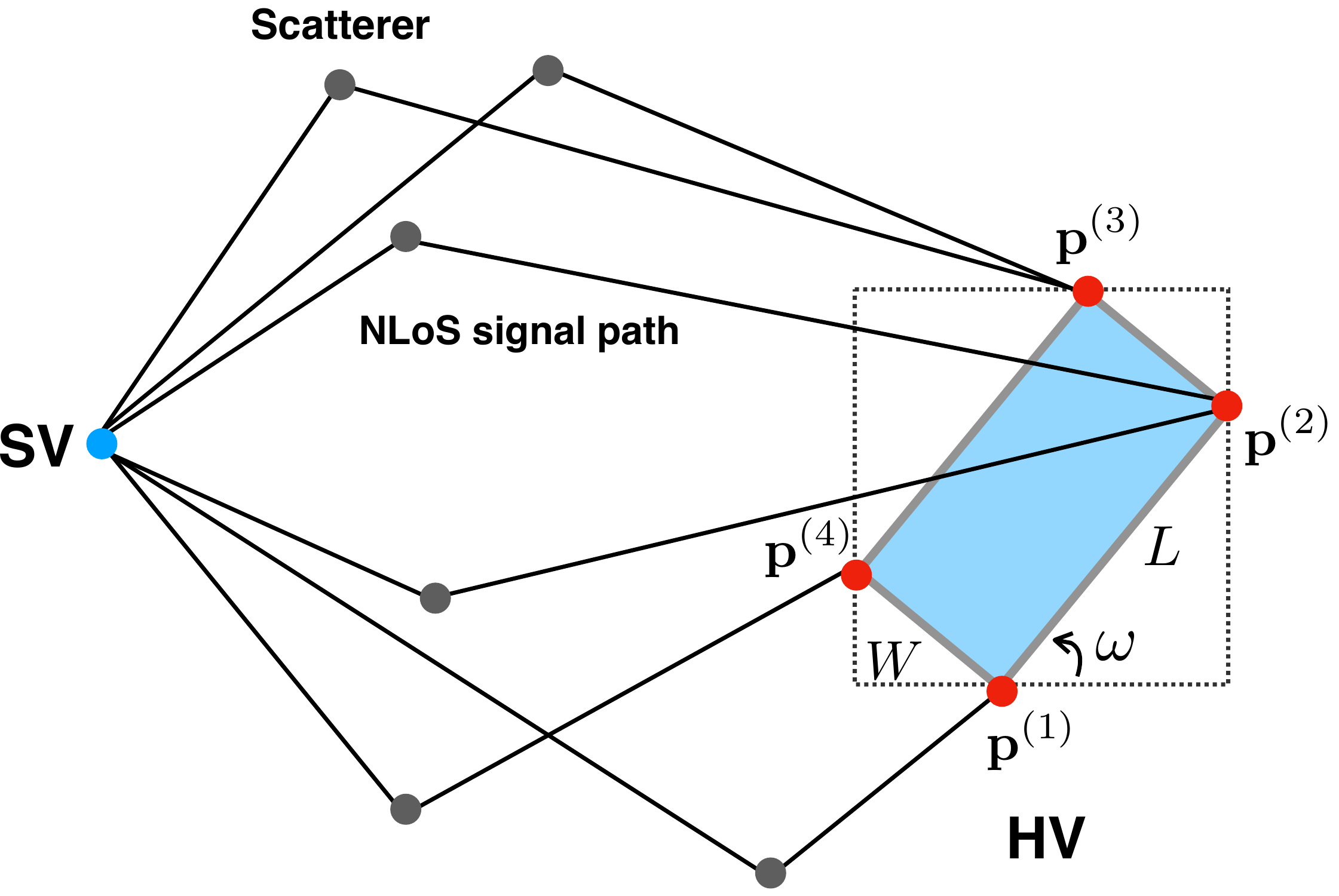}
\caption{Rectangular configuration of  $4$-cluster arrays at HV.}\label{fourAnteLS}
\vspace{-20pt}
\end{figure}

\section{Sensing Hidden Vehicles with Multi-Cluster  Arrays}\label{sec:fourPositioning}
Consider the case that the HV arrays consists of four antenna clusters located at the vertices of a rectangle with length $L$ and width $W$ (see Fig.~\ref{fourAnteLS}). The vertex locations are represented as $\{\bp^{(k)} = (x^{(k)}, y^{(k)})^{\mathrm{T}}\}_{k=1}^4$. Recall that the SV can differentiate the origin from which signal is transmitted due to the usage of different orthogonal waveform set for each array. Let each path be ordered based on HV arrays' index such that $\mathcal{P} = \{\mathcal{P} ^{(1)}, \mathcal{P} ^{(2)}, \mathcal{P} ^{(3)}, \mathcal{P} ^{(4)}\}$ where
$\mathcal{P} ^{(k)}$ represents the set of received signals from the $k$-th array.
Note that the vertices determines  the shape and their centroid of HV location. Therefore, the \textbf{sensing problem} is represented as
\begin{equation}\label{Eq:ProbForm4}
\begin{aligned}
\bigcup\nolimits_{k =1 }^4 \bigcup\nolimits_{p \in \mathcal{P}^{(k)} } \{\theta_p, \phi_p, \lambda_p\} \Rightarrow \{\{\bp^{(k)}\}_{k=1}^4,  \omega\}.
\end{aligned}
\end{equation}
Next, we present a sensing technique exploiting prior knowledge of the HV $4$-cluster arrays' configuration, which   is more efficient than separately estimating the four positions  $\{\bp^{(k)}\}_{k=1}^4$ using the technique in the preceding section.

\subsection{Sensing Feasibility Condition}
 Assume that $\mathcal{P}^{(1)}$ is not empty and $1\in \mathcal{P}^{(1)}$ without loss of generality.
Based on  the rectangular configuration  of $\{\mathbf{p}^{(k)}\}_{k=1}^4$ (see Fig.~\ref{fourAnteLS}),
a  system of equations is formed:
\begin{align}\label{FGrelation_4MIMO}\tag{P3}\!\!\!\!
\begin{cases}
\nu_{p} \cos(\theta_{p}) - (d_{p} - \nu_{p}) \cos(\varphi_{p} + \omega)+\eta_p(\omega,L, W) \\= \nu_{1} \cos(\theta_{1}) - (d_{1} - \nu_{1}) \cos(\varphi_{1} + \omega),  \\
\nu_{p} \sin(\theta_{p}) - (d_{p} - \nu_{p}) \sin(\varphi_{p} + \omega)+\zeta_p(\omega, L, W ) \\ =\nu_{1} \sin(\theta_{1}) - (d_{1} - \nu_{1}) \sin(\varphi_{1} + \omega),
\end{cases}
\end{align}
where
\begin{align}\label{funcLW1}
\eta_p(\omega, L, W)= \l\{
\begin{aligned}
&0, && \textrm{$p\in\mathcal{P}^{(1)}$}\\
&L\cdot \cos(\omega), && \textrm{$p\in\mathcal{P}^{(2)}$}\\
&L\cdot \cos(\omega)-W\cdot \sin(\omega), && \textrm{$p\in\mathcal{P}^{(3)}$}\\
&-W\cdot\sin(\omega), && \textrm{$p\in\mathcal{P}^{(4)}$}
\end{aligned}
\r.
\end{align}
and $\zeta_p(\omega, L, W)$ is obtained via replacing all $\cos$ and $\sin$ in $\eqref{funcLW1}$ with $\sin$ and $-\cos$, respectively. Recall $P=|\mathcal{P}|=\sum_{k=1}^4 |\mathcal{P}^{(k)}|$. Compared with \ref{FGrelation}, the number of equations in \ref{FGrelation_4MIMO} is the same as $2(P-1)$ while the number of unknowns increases from $P+2$ to $P+4$ because $L$ and $W$ are also unknown. Consequently, \ref{FGrelation_4MIMO} has a unique solution when $2(P-1)\geq P+4$.

\begin{proposition}[Sensing feasibility condition]\label{pro:minNumPath_4MIMO}
\emph{To sense the position, shape, and orientation of a HV with $4$-cluster arrays, at least six paths are required: $P\geq 6$.}
\end{proposition}

\begin{remark}[Advantage of array-configuration   knowledge]\emph{The separate positioning  of individual HV $4$-cluster arrays requires at least $16$ NLoS paths (see Proposition \ref{pro:minNumPath}). On the other hand, the prior knowledge of rectangular configuration  of antenna clusters leads to  the relation between  their locations, reducing the number of required paths for sensing.}
\end{remark}

\subsection{Hidden Vehicle Sensing}
Consider the case that noise is neglected. \ref{ax=b} is rewritten to the following matrix form:
\begin{align}\label{ax=b_4MIMO}\tag{P4}
\hat{\mathbf{A}}(\omega) \hat{\mathbf{z}}= {\mathbf{B}}(\omega),
\end{align}
where $\hat{\mathbf{z}} =(\mathbf{v}, d_1, L, W)^{\mathrm{T}} \in \mathds{R}^{(P+3)\times 1}$ with $\mathbf{v}$
 following the index ordering of $\mathcal{P}$, and ${\mathbf{B}}(\omega)$ is given in \eqref{matrixB}. For matrix $\hat{\mathbf{A}}(\omega)$, we have
\begin{align}\label{matrixA_4MIMO}
\hat{\mathbf{A}}(\omega) &=
\begin{bmatrix}
  \mathbf{A}(\omega)&\mathbf{L}(\omega)&\mathbf{W}(\omega)
\end{bmatrix}
\in \mathds{R}^{2(P-1) \times (P+3)}.
\end{align}
Here, $\mathbf{A}(\omega)$ is specified in \eqref{matrixA} and $\mathbf{L}(\omega)\in \mathds{R}^{2(P-1) \times 1}$ is given as $[\mathbf{L}^{(\cos)}(\omega), \mathbf{L}^{(\sin)}(\omega)]^{\mathrm{T}}$ where
\begin{align}
\mathbf{L}^{(\cos)}(\omega) &= [\underbrace{0,\cdots,0}_{|\mathcal{P}^{(1)}|-1}, \underbrace{-\cos(\omega),\cdots, -\cos(\omega)}_{|\mathcal{P}^{(2)}|+|\mathcal{P}^{(3)}|}, \underbrace{0,\cdots,0}_{|\mathcal{P}^{(4)}|}  ]^{\mathrm{T}}, \nn
\end{align}
and $\mathbf{L}^{(\sin)}(\omega)$ is obtained by replacing all $\cos(\omega)$ in $\mathbf{L}^{(\cos)}(\omega)$ with $\sin(\omega)$. Similarly, $\mathbf{W}(\omega) $ is given as $ [\mathbf{W}^{(\sin)}(\omega), \mathbf{W}^{(\cos)}(\omega)]^{\mathrm{T}}$ where
\begin{align}
\mathbf{W}^{(\sin)}(\omega) &= [\underbrace{0,\cdots,0}_{|\mathcal{P}^{(1)}|+|\mathcal{P}^{(2)}|-1}, \underbrace{\sin(\omega),\cdots, \sin(\omega)}_{|\mathcal{P}^{(3)}|+|\mathcal{P}^{(4)}|}]^{\mathrm{T}}, \nn
\end{align}
and $\mathbf{W}^{(\cos)}(\omega)$ is obtained by replacing all $\sin$ in $\mathbf{W}^{(\sin)}(\omega)$ with $-\cos$.

\noindent \underline{1) Computing $\omega^*$:}
Noting that \ref{ax=b_4MIMO} is over-determined when $P\geq 6$, the resultant discriminant of the orientation $\omega$ is similar to Proposition~\ref{pro:discriminant} and given as follows.

\begin{proposition}[Discriminant of orientation]\label{pro:discriminantMultiCluster}
\emph{With $P \geq 6$, the unique $\omega^*$ exists when $\hat{\mathbf{B}}(\omega^*)$ is orthogonal to the null column space of $\hat{\mathbf{A}}(\omega^*)$ denoted by $\mathsf{null}(\hat{\mathbf{A}}(\omega^*)^{\mathrm{T}})\in \mathds{R}^{2(P-1)\times (P+1)}$:
\begin{align}\label{Eq:OrthogonalNullSpaceMultiCluster}
\mathsf{null}(\hat{\mathbf{A}}(\omega^*)^{\mathrm{T}})^{\mathrm{T}}\hat{\mathbf{B}}(\omega^*)=\mathbf{0}.
\end{align}}
\end{proposition}
\noindent 
Given  this discriminant, a simple 1D search can be performed over the range $[0, 2\pi]$ to find  $\omega^*$.

\noindent \underline{2) Computing $\{\bp^{(k)}\}_{k=1}^4$:}
Given the $\omega^*$, \ref{ax=b_4MIMO} can be  solved by
\begin{align}\label{LSresult4MIMO}
\hat{\mathbf{z}}^* = \l[ \hat{\mathbf{A}}(\omega^*)^{\mathrm{T}} \hat{\mathbf{A}}(\omega^*)  \r]^{-1} \hat{\mathbf{A}}(\omega^*)^{\mathrm{T}} \hat{\mathbf{B}}(\omega^*).
\end{align}
HV arrays' positions $\{\bp^{(k)}\}_{k=1}^4$ can be  computed  by substituting \eqref{Eq:OrthogonalNullSpaceMultiCluster} and \eqref{LSresult4MIMO} into \eqref{Eq:TDoAandPropagationDist} and \ref{FGrelation_4MIMO}.

Extending the technique  to the case with noise is omitted for brevity because it is straightforward by modifying \eqref{Eq:OrthogonalNullSpaceMultiCluster} to a minimization problem as in Sec.~\ref{subsection:NoiseHVSensing}.

\begin{figure}[t]
\vspace{-10pt}
\centering
\includegraphics[width=7.5cm]{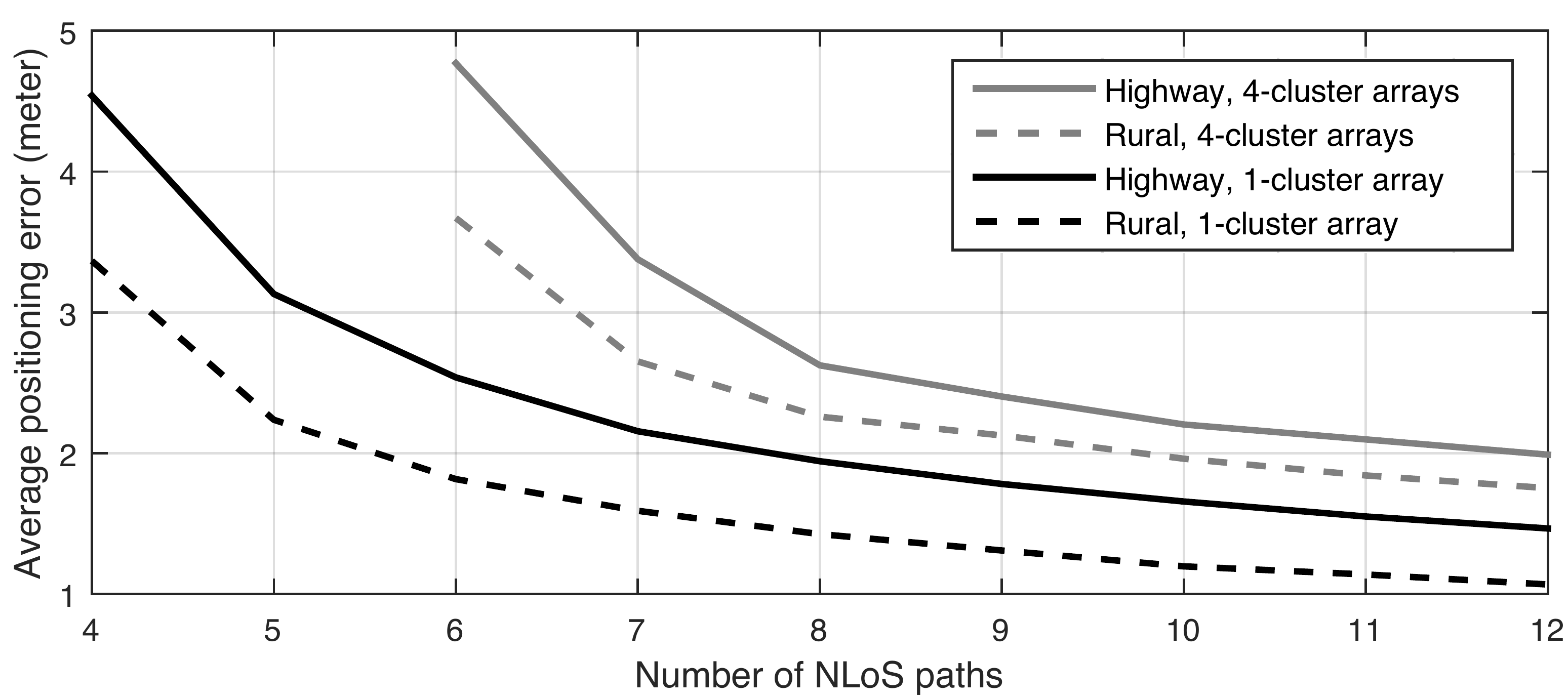}
\caption{Number of NLoS paths versus average positioning error.}\label{Sim:NumPaths}
\vspace{-10pt}
\end{figure}

\begin{figure}[t]
\vspace{-0pt}
\centering
\includegraphics[width=7.5cm]{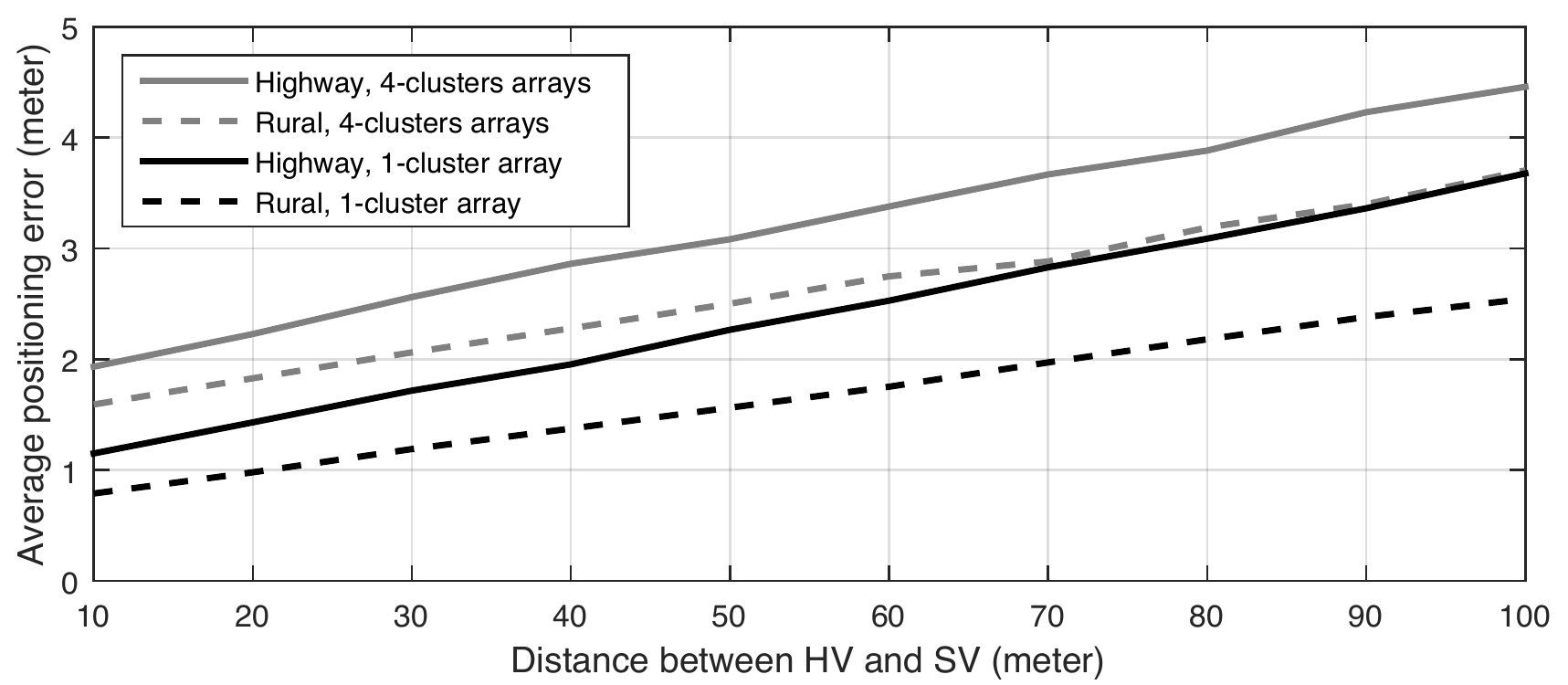}
\caption{ SV-HV distance versus average positioning error.}\label{Sim:Dis}
\vspace{-10pt}
\end{figure}

\section{Simulation Results}
The performance of the proposed technique is validated via realistic simulation. The performance metric for measuring positioning accuracy is defined as the average Euclidean squared distance of estimated arrays' positions to their true locations: $\frac{1}{4} \sum_{k=1}^{4}\| \bp^{*(k)} - \bp^{(k)}\|^2$, named \emph{average positioning error}. We adopt the geometry-based stochastic channel model given in \cite{karedal2009geometry} for modelling the practical scatterers distribution and V2V propagation channels, which has been validated by real measurement data. Two scenarios, highway and rural, are considered by following the  settings in \cite[Table 1]{karedal2009geometry}. We set $f_c = 5.9$ GHz, $B_s = 100$ MHz, $M_r = M_t = 20$, the per-antenna transmission power  is $23$ dBm. The size of HV is $L\times W = 3\times 6~\textrm{m}^2$ and distance between SV and HV is $50~\textrm{m}$.

Fig.~\ref{Sim:NumPaths} shows the curves of average positioning error versus the number of NLoS paths $P$ received at SV. It is observed that positioning via $1$ and $4$-cluster arrays are feasible when the $P\geq4$ and $P\geq6$, respectively, and receiving more paths can dramatically decrease the positioning error. The error for the $4$-cluster arrays is much larger. This is because more clusters results in more  noise, which leads to noisy estimations of AoA/AoD/ToAs within signal detection procedure. Also, compared with $1$-cluster array, two more unknown parameters need to be jointly estimated in the  case of $4$-cluster arrays, which impacts the positioning performance. Moreover, the positioning accuracy in the rural scenario is better than that in highway scenario. The reason is that the signal propagation loss in highway scenario is higher than that in rural scenario since the distance between vehicle and scatterers can be large, which adds the difficulty for signal detections.

In Fig.~\ref{Sim:Dis}, the distance between SV and HV versus average positioning error is plotted. It is shown that the positioning error increases when SV-HV distance keeps increasing because the accuracy of signal detection reduces when SV-HV distance becomes larger since higher signal propagation loss. The positioning accuracy in rural scenario is higher than that in highway. The reason is that more paths can be received at SV in rural case due to the denser scatterers exists, resulting in higher positioning accuracy as Fig.~\ref{Sim:NumPaths} displays. Moreover, the error gap between highway and rural cases increases with SV-HV distance. This is because, as the SV-HV distance increases, the power of received signals in highway is weaker than those in rural due to larger propagation loss, leading to inaccurate signal detections.

\section{Conclusion Remarks}
A novel and efficient  technique has been  proposed for sensing hidden vehicles. Presently, we are extending  the technique to  the case where the SV has no knowledge of waveform assignments to different HV arrays, and  to 3D propagation.


\end{document}

%% file: header.tex
\newtheorem{proposition}{Proposition}

\newtheorem{remark}{\bf Remark}
\def\phi{\varphi}

\def\l{\left}
\def\r{\right}
\def\({\left(}
\def\){\right)}

\def\bp{{\mathbf{p}}}

\def\b0{{\mathbf{0}}}

\newcommand{\nn}{\nonumber}

%% file: Vehicle_Draft_Conference_v5.bbl
\begin{thebibliography}{10}

\bibitem{alam2013cooperative}
N.~Alam and A.~Dempster, ``Cooperative positioning for vehicular networks:
  Facts and future,'' {\em IEEE Trans. Intell. Transp. Syst.}, vol.~14,
  pp.~1708--1717, Dec. 2013.

\bibitem{choi2016millimeter}
J.~Choi and et~al., ``Millimeter-wave vehicular communication to support
  massive automotive sensing,'' {\em IEEE Commun. Mag.}, vol.~54, pp.~160--167,
  Dec. 2016.

\bibitem{rossi2014spatial}
M.~Rossi, A.~Haimovich, and Y.~Eldar, ``Spatial compressive sensing for mimo
  radar,'' {\em IEEE Trans. Sig. Proc.}, vol.~62, pp.~419--430, Jan. 2014.

\bibitem{schwarz2010lidar}
B.~Schwarz, ``Lidar: Mapping the world in 3d,'' {\em Nature Photonics}, vol.~4,
  pp.~429--430, Jul. 2010.

\bibitem{gezici2005localization}
S.~Gezici and et~al, ``Localization via ultra-wideband radios: a look at
  positioning aspects for future sensor networks,'' {\em IEEE Signal Proc.
  Mag.}, vol.~22, pp.~70--84, Jul. 2005.

\bibitem{shin2002comparisons}
D.-H. Shin and T.-K. Sung, ``Comparisons of error characteristics between toa
  and tdoa positioning,'' {\em IEEE Trans Aerospace Elec. Systems}, vol.~38,
  pp.~307--311, Jan. 2002.

\bibitem{miao2007positioning}
H.~Miao, K.~Yu, and M.~J. Juntti, ``Positioning for {NLOS} propagation:
  Algorithm derivations and {C}ramer-{R}ao bounds,'' {\em IEEE Trans. Veh.
  Tech.}, vol.~56, pp.~2568--2580, Sep. 2007.

\bibitem{shahmansoori2018position}
A.~Shahmansoori and et~al, ``Position and orientation estimation through
  millimeter-wave {MIMO} in 5{G} systems,'' {\em IEEE Trans. Wireless Commun.},
  vol.~17, pp.~1822--1835, Mar. 2018.

\bibitem{Cheng2013}
L.~Cheng, D.~Stancil, and F.~Bai, ``A roadside scattering model for the
  vehicle-to-vehicle communication channel,'' {\em IEEE J. Sel. Areas Commun.},
  vol.~31, pp.~449--459, Sep. 2013.

\bibitem{cui2016vehicle}
X.~Cui, T.~Gulliver, J.~Li, and H.~Zhang, ``Vehicle positioning using 5{G}
  {M}illimeter-wave systems,'' {\em IEEE Access}, vol.~4, pp.~6964--6973, Oct.
  2016.

\bibitem{mathews1994eigenstructure}
C.~Mathews and M.~Zoltowski, ``Eigenstructure techniques for 2-{D} angle
  estimation with uniform circular arrays,'' {\em IEEE Trans. Sig. Proc.},
  vol.~42, pp.~2395--2407, Sep. 1994.

\bibitem{karedal2009geometry}
J.~Karedal and et~al., ``A geometry-based stochastic {MIMO} model for
  vehicle-to-vehicle communications,'' {\em IEEE Trans. Wireless Commun.},
  vol.~8, pp.~3646--3657, Jul. 2009.

\end{thebibliography}
